\documentclass[amsmath,amssymb,prb,superscriptaddress, twocolumn]{revtex4}

\usepackage{graphicx}
\usepackage{amssymb,amsfonts,amsmath}

\newcommand{\unit}[1]{\ensuremath{\, \mathrm{#1}}}
\DeclareMathOperator{\im}{Im}
\DeclareMathOperator{\tr}{Tr}

\begin{document}



\title{Strong quantum memory at resonant Fermi edges revealed by shot noise}

\author{N. Ubbelohde}
\affiliation{Institut f\"ur Festk\"orperphysik, Leibniz Universit\"at Hannover, Appelstra{\ss}e 2, D-30167 Hannover, Germany}

\author{K. Roszak}
\affiliation{Department of Condensed Matter Physics,
             Faculty of Mathematics and Physics, Charles University,
             CZ-12116 Prague, Czech Republic}
\affiliation{Institute of Physics, Wroc{\l}aw University of Technology, PL-50370 Wroc{\l}aw, Poland}

\author{F. Hohls}
\affiliation{Institut f\"ur Festk\"orperphysik, Leibniz Universit\"at Hannover, Appelstra{\ss}e 2, D-30167 Hannover, Germany}
\affiliation{Physikalisch-Technische Bundesanstalt, D-38116 Braunschweig, Germany\\
{$^{*}$\rm Correspondence and requests for materials should be addressed to T.N. (tno@karlov.mff.cuni.cz)}}

\author{N. Maire}
\affiliation{Institut f\"ur Festk\"orperphysik, Leibniz Universit\"at Hannover, Appelstra{\ss}e 2, D-30167 Hannover, Germany}

\author{R. J. Haug}
\affiliation{Institut f\"ur Festk\"orperphysik, Leibniz Universit\"at Hannover, Appelstra{\ss}e 2, D-30167 Hannover, Germany}

\author{T. Novotn\'{y}$^{*,}$}
\affiliation{Department of Condensed Matter Physics,
             Faculty of Mathematics and Physics, Charles University,
             CZ-12116 Prague, Czech Republic}


\begin{abstract}  
Studies of non-equilibrium current fluctuations enable assessing correlations involved in quantum transport through nanoscale conductors. They provide additional information to the mean current on charge statistics and the presence of coherence, dissipation, disorder, or entanglement. Shot noise, being a temporal integral of the current autocorrelation function, reveals dynamical information. In particular, it detects presence of non-Markovian dynamics, i.e., memory, within open systems, which has been subject of many current theoretical studies. We report on low-temperature shot noise measurements of electronic transport through InAs quantum dots in the Fermi-edge singularity regime and show that  it exhibits strong memory effects caused by quantum correlations between the dot and fermionic reservoirs. Our work, apart from addressing noise in archetypical strongly correlated system of prime interest, discloses generic quantum dynamical mechanism occurring at interacting resonant Fermi edges.
\end{abstract}


\date{\today}
\maketitle

Non-equilibrium electronic shot noise is a powerful diagnostic tool revealing properties of mesoscopic systems inaccessible by the mean current measurements.\cite{PhysicsToday,Nazarov:book,Blanter} For example, recent noise measurements in the Kondo impurities\cite{Glattli:NatPhys09,Yamauchi:PRL11} have brought new insights into strongly correlated transport. Shot noise is sensitive to the presence of non-Markovian dynamics\cite{Flindt:PRL08} intensively studied in broad context ranging from photosynthesis to quantum information.\cite{Fleming:Science07, Erez:Nature08, Eisert:PRL08, Breuer:PRL09, Plenio:PRL10, Piilo:NatPhys11} However, most theoretical proposals as well as the newest quantum-optical experimental study\cite{Piilo:NatPhys11} rely on extensive engineering and control of the system and/or environment (bath) and a clear observation and identification of the quantum memory effects in ``natural'', i.e., routinely fabricated solid-state systems has not been reported yet. 

In resonant tunneling, which is ubiquitous in quantum electronic transport, the charge dynamics of the resonant level can be described by a simple Markovian master equation\cite{Blanter} as long as the relaxation time related to the inverse of the transfer rates is long compared to the characteristic memory time of the fermionic bath (leads) given by the inverse temperature and/or the detuning of the level from the chemical potentials of the leads. Comparable time scales for system relaxation and bath memory break down the conventional description for low-temperature on-resonance transport  and indicate\cite{Hanggi:PRL05} strong non-Markovian features which, together with many body-interactions typical for small nanostructures, influence the low-temperature width of the resonant steps in the current-voltage characteristics,\cite{Kubala:PRB06} the decay of the level occupations,\cite{Rabani:PRB11} or the noise.\cite{Flindt:PRL08} For the noise, significant deviations from the conventional master equation description have already been observed,\cite{Maire:PRB07} although their origin has not been identified.

\begin{center}
\begin{figure*}
\includegraphics[width=\textwidth]{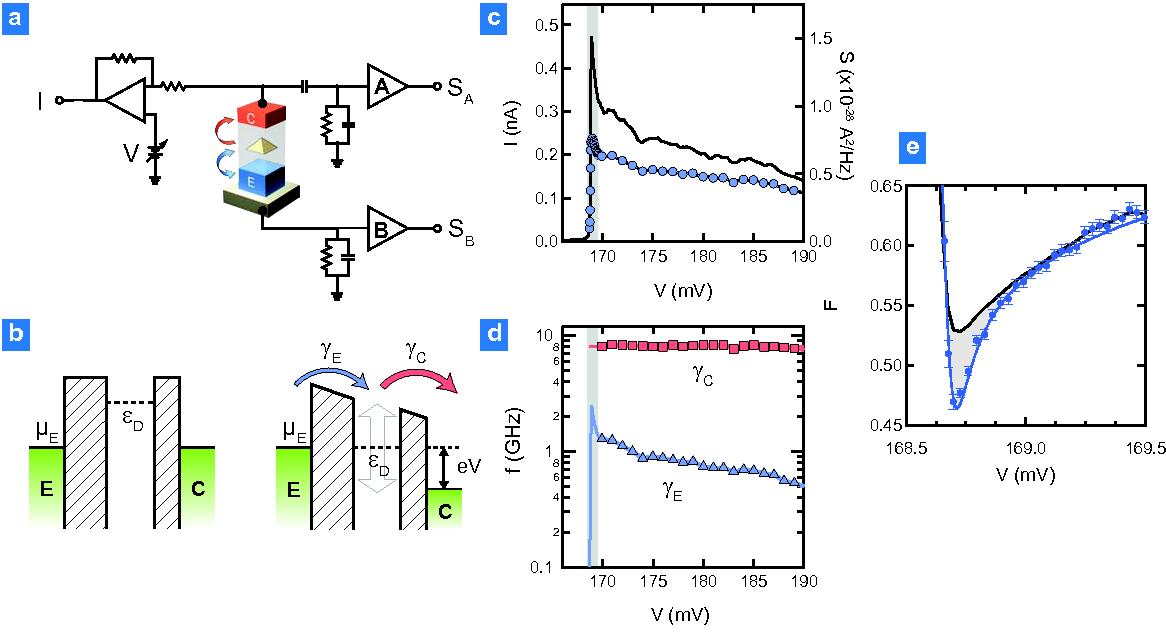}
\caption{{\bf Shot noise power measurement around the Fermi edge singularity.} (a) Simplified schematic of the studied device consisting of an InAs QD (pyramid) between emitter (E, blue) and collector (C, red) and the equivalent detection circuit. (b) Energy levels of the dot and leads. Left: zero applied bias with unoccupied dot level high above the lead chemical potential. Right: threshold bias when the dot level aligns with the emitter chemical potential and the current strongly enhanced by the Fermi edge singularity starts flowing. (c) Mean current $I$ (solid line) and shot noise power $S$ (symbols) as functions of applied voltage at $T = 70\unit{mK}$ and $B = 0\unit{T}$. (d) Energy dependence of $\gamma_E$ and $\gamma_C$ in the off-resonant regime as determined from the measured current and shot noise (symbols outside the shaded region) and around the resonance with $\gamma_{C}$ fixed and $\gamma_{E}$ calculated from the mean current (lines within the shaded region). (e) Fano factor around the resonant edge (corresponding to the shaded range of (c) and (d)). Experimental values supplemented with their estimated errors (details in the Methods section) are contrasted with the Markovian approximation based on tunnel rate values from (d) (black line) and the full non-Markovian theory (blue line).}\label{fig:fig1}
\end{figure*}
\end{center}

In this work we present new experimental results on the low-temperature noise measurements in the Fermi-edge singularity (FES) regime\cite{Matveev:PRB92, Geim:PRL94, Hapke-Wurst:PRB00,Frahm:PRB06,Maire:PRB07}  together with a theoretical analysis clearly revealing the presence of strong quantum memory around the edge. The noise-around-the-edge puzzle\cite{Maire:PRB07} is briefly introduced in Fig.~\ref{fig:fig1}e, where the measured points are contrasted with the standard Markovian theory\cite{Blanter} (black line) showing large deviations of $\sim15\%$ in the Fano factor $F\equiv S/2eI$ (a convenient dimensionless measure of the shot noise $S$), far beyond the experimental uncertainty. Moreover, the measured dip breaks the Markovian lower bound\cite{Blanter} of $1/2$, which is a clear witness  of strong memory. The blue line, nicely coinciding with the data, is our new theory accounting for the memory effects.

The FES, a paradigmatic exactly solvable many-body problem,\cite{Nozieres:PR69,Ohtaka:RMP90} which originates from the Coulomb interaction of conduction electrons with those on a localised discrete level represented by core shell electrons or quantum dot (QD) levels, was first predicted in the X-ray spectra of metals,\cite{Mahan:PR67} but its signatures are observed also in resonant tunneling set-ups as a (truncated) power-law singularity  of the mean current $I$ around, e.g., the emitter Fermi energy.\cite{Matveev:PRB92, Geim:PRL94,Hapke-Wurst:PRB00,Frahm:PRB06} The interacting resonant level model describing the FES transport setup has served recently as an important benchmark for novel quantum transport techniques\cite{Andrei:PRL06, Doyon:PRL07, Schmitteckert:PRL08} including the noise calculation\cite{Schmitteckert:PRL10} at its exactly solvable self-dual point (different from our experimental regime).
\newline

{\noindent\large\bf Results\\}
We first describe cross-correlation measurements of current shot noise in self-assembled InAs QDs in the FES regime. The experimental set-up is depicted in Fig.~\ref{fig:fig1}a and explained in more detail in the Methods section. At zero bias voltage the ground state energy level $\varepsilon_D$ of the InAs dots lies far above the emitter Fermi energy $\mu_{E}$ (see Fig.~\ref{fig:fig1}b, left). Therefore, a large threshold voltage bias $V_{\rm th}\approx 170 \unit{mV}$ applied to the collector lead is required to shift $\varepsilon_D$ to resonance with the emitter Fermi energy by electrostatic gating with the leverage factor $\eta=\frac{\mu_E-\varepsilon_D}{e(V-V_{\rm th})}$ giving the fraction of the bias voltage dropped at the emitter-dot junction (see Fig.~\ref{fig:fig1}b, right). On resonance, the tunneling current sets in and displays a sharp peak (shaded part of Fig.~\ref{fig:fig1}c, solid line) due to the Fermi edge singularity caused by the Coulomb interaction of the occupied dot level with the electrons in the emitter lead (there is no relevant interaction with the collector due to the asymmetry of the setup). Further increase of the bias causes a decrease of the current due to the decrease of the emitter rate induced by the three-dimensional density of states (DOS) in the emitter.\cite{Hapke-Wurst:PRB00} Together with the current-voltage characteristics on a large voltage scale, i.e., far around the edge, Fig.~\ref{fig:fig1}c shows the measured shot noise power $S$ (symbols).

Far enough from the edge, i.e., outside of the shaded region of Fig.~\ref{fig:fig1}c, we can use the standard Markovian master equation and evaluate the emitter $\gamma_E$ and collector $\gamma_C$ tunneling rates, Fig.~\ref{fig:fig1}d, from formulas \cite{Blanter} $I = 2e\gamma_E \gamma_C / (2\gamma_E+\gamma_C)$ and $F = 1-4\gamma_E \gamma_C/(2\gamma_E+\gamma_C)^2$ (excluded double occupancy due to strong onsite Coulomb interaction implies usage of $2\gamma_E$ instead of just $\gamma_E$ as for a noninteracting resonant level\cite{Gurvitz:PRB96}). While the collector rate $\gamma_C\doteq 8\cdot 10^{9}s^{-1}$ is basically constant, $\gamma_E$ reflects the energy dependence of the emitter DOS \cite{Nauen:PRB04} and exhibits an expected asymmetry of the tunneling barriers with $\gamma_E/\gamma_C$ ranging between 0.06 and 0.22. Plausibly assuming constant $\gamma_{C}$ throughout the resonance we can analogously to Ref.~\onlinecite{Maire:PRB07} extrapolate the $\gamma_{E}$ to the resonance (shaded) region in Fig.~\ref{fig:fig1}d (solid lines) from the  expression for the current. The resulting {\em Markovian prediction} based on these extrapolated rates (black curve in Fig.~\ref{fig:fig1}e) clearly exhibits substantial deviations from the measurement inexplicable by experimental errors. Obviously, the resonant transport regime calls for a radically new theoretical understanding.

{\noindent\large\bf Discussion\\}
Using the procedure briefly described in the Methods section for $B=0$, we arrive at a non-Markovian generalised
master equation (GME) for the occupations of the resonant level, where
$p_{1}(t)$ is the probability that the level is occupied by an electron,
while $p_{0}(t)=1-p_{1}(t)$ denotes the probability of the dot being empty,
\begin{equation}\label{GME}
\frac{dp_0(t)}{dt}=\gamma_C p_1(t)+\int_0^t dt' [\gamma_{E}^{b}(t')p_1(t-t')-2\gamma_{E}^{f}(t')p_0(t-t')].
\end{equation}
The expressions ($\hbar=1$) for the forward/backward non-Markovian electron transfer rates across the QD/emitter-lead interface
$\gamma_{E}^{f/b}(t)=2\Re\!\left[e^{-(\frac{\gamma_C}{2}-i\varepsilon_D)t} G_{0/1}(t)\right]$
involve standard FES Green's functions, whose evaluation is a known result of the FES theory.\cite{Nozieres:PR69,Ohtaka:RMP90,Levitov:PRL05}
This results in the explicit form of the rates entering equation \eqref{GME} ---
in the Laplace space they read
$\gamma_E^{f}(z;\Delta)\propto-\Im\left[\left(\frac{-i}{2\pi k_B T}\right)^{\alpha}
B\left(\frac{1-\alpha}{2}+\frac{z+\gamma_C(1+i\Delta)/2}{2\pi k_B T},\alpha\right)\right]$
and $\gamma_E^b(z;\Delta)=\gamma_E^{f }(z;-\Delta)$,
where $\Delta\equiv\tfrac{\mu_E-\varepsilon_D}{\gamma_C/2}=\tfrac{e\eta(V-V_{\rm th})}{\gamma_C/2}$ is the dimensionless energy/voltage distance from the resonant edge, $\alpha$ is the FES critical exponent,
and $B(x,y)$ denotes the beta-function.  In the zero-temperature limit the formula simplifies to 
$\gamma_E^{f}(z;\Delta)\propto -\Im\left[\left(\frac{-i}{z+\gamma_C(1+i\Delta)/2}\right)^{\alpha} \right]$.  When the counting field $\chi$ at the emitter junction is included \cite{Flindt:PRL08,Flindt:PRB10}
the GME memory kernel corresponding to equation \eqref{GME} is of the form
\begin{equation}
\label{kernel}
\mathcal{W}(\chi,z;\Delta)=
\begin{pmatrix}
-2\gamma_E^{f}(z;\Delta) &\gamma_E^b(z;\Delta)e^{-\chi} + \gamma_C \\
2\gamma_E^{f}(z;\Delta)e^{\chi} &-\gamma_E^b(z;\Delta) -\gamma_C
\end{pmatrix}.
\end{equation}

\begin{figure}[h!]
\includegraphics[width=0.45\textwidth]{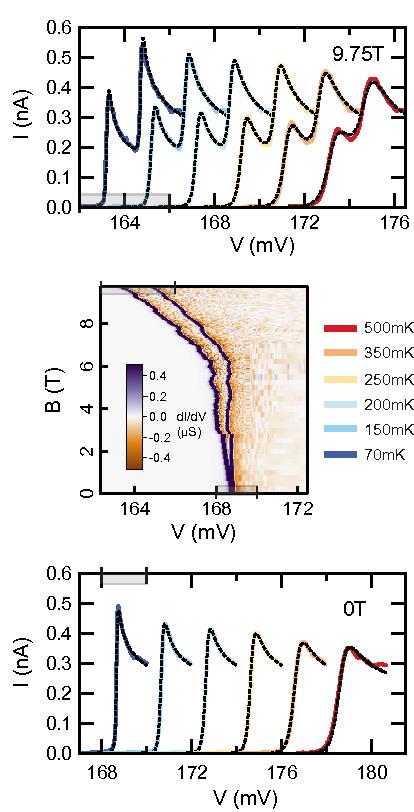}
\caption{{\bf Mean current fits for FES in magnetic field.}
Top (bottom) panel: $I$-$V$ curves for different temperatures (colour-coded as shown by the middle panel and horizontally shifted with respect to each other for clarity)
at $B=9.75 \unit{T}$ ($B=0$). Experimental data are plotted in colour and the black dashed curves are the theoretical fits for a {\em unique} set of 6 (4) parameters at all temperatures. Middle panel: Differential conductance $dI/dV$ as a function of the bias voltage and magnetic field at the lowest temperature $T=70 \unit{mK}$. The Zeeman splitting of the edge is clearly visible. The top and bottom panels depict the cuts along the corresponding borders of the middle panel; the shaded stripes in the panels indicate matching ranges for the lowest temperature curves.}
\label{fig:fig2}
\end{figure}

Using the standard procedure for the cumulant evaluation \cite{Flindt:PRL08, Flindt:PRB10}
on this memory kernel, we get, using the abbreviations $\gamma_E^{f/b}\equiv\gamma_E^{f/b}(0;\Delta)$ and $\gamma_E^{f/b\prime}\equiv d\gamma_E^{f/b}(z;\Delta)/dz|_{z=0}$ , the formulas for the mean current $I=e\tfrac{2\gamma_C\gamma_E^{f}}{\gamma_C+2\gamma_E^{f}+\gamma_E^b}$
and for the non-Markovian Fano factor
\begin{equation}\label{Fanofactor}
F=1-4\frac{\gamma_C\gamma_E^{f}}{(\gamma_C+2\gamma_E^{f}+\gamma_E^b)^2}+4\gamma_C\frac{(\gamma_C+\gamma_E^b)\gamma_E^{f\prime}-\gamma_E^{f}
\gamma_E^{b\prime}}{(\gamma_C+2\gamma_E^{f}+\gamma_E^b)^2}.
\end{equation}
The last term in the Fano factor, proportional to the derivatives, constitutes the non-Markovian correction. Well above the edge, where both the back-flow $\gamma_{E}^{b}$ and the non-Markovian features can be neglected, we recover the standard master equation result (with $\gamma_E\equiv\gamma_E^{f}$).

\begin{center}
\begin{figure*}
\includegraphics[width=\textwidth]{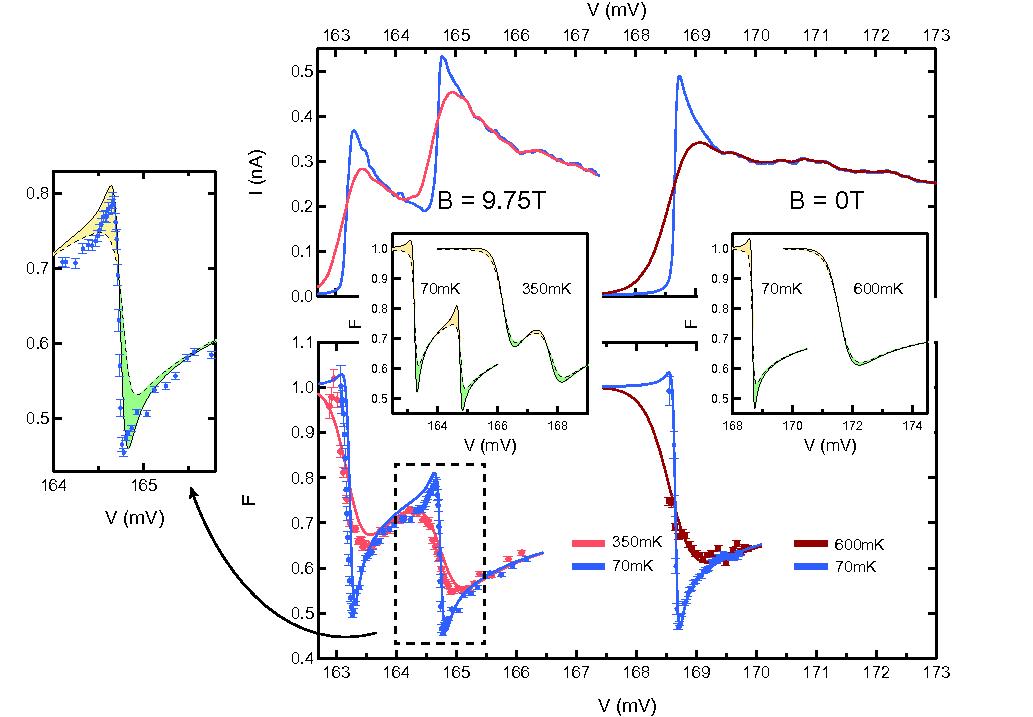}
\caption{{\bf Memory effects on the FES noise.} Top (bottom) panel: mean current (Fano factor) for two values of the magnetic field specified in the top panel and various temperatures shown in the bottom panel. Measured Fano factor with the estimated error-bars (explained in the Methods section) is compared to theoretical predictions based on parameters obtained from the fits of Fig.~\ref{fig:fig2}. Insets: Comparison of non-Markovian theory (solid lines) with the Markovian approximation (dashed lines) for corresponding magnetic fields (individual insets) and temperatures (curves within insets; horizontally displaced for clarity). Differences between the two curves are highlighted by colours according to their sign. Left detail: zoom onto the low-temperature Fano factor curve in the region around the upper Fermi edge shown by the dashed rectangle. Measured data with their error-bars are supplemented with both the non-Markovian as well as Markovian predictions in the spirit of the insets.}\label{fig:fig3}
\end{figure*}
\end{center}

In the lowest order in $\gamma_E^{f,b}/\gamma_C$ we can write
$F\approx 1- 4\gamma_E^{f}/\gamma_C+4\gamma_E^{f\prime}$, with magnitude $|\gamma_E^{f\prime}/\gamma_{E}^{f}|\approx \big[\max( k_{B}T, \gamma_{C}, \gamma_{C}\Delta/2)\big]^{-1}$.
This implies that for low temperatures
$k_{B}T\lesssim\gamma_{C}$ and close to the edge
$|\Delta|\lesssim 2$ the non-Markovian correction is governed by the collector rate $\gamma_{C}$ and, thus,
it is of the {\em same order} as the Markovian correction $- 4\gamma_E^{f}/\gamma_C$ to the Poissonian noise
(with $F=1$) due to correlations caused by sequential occupying and emptying of the QD. Being of quantum origin, it vanishes fast with increasing temperature $k_{B}T\gtrsim\gamma_{C}$, which kills quantum correlations between the dot and leads responsible for the memory effects. Moreover, it generically assumes both signs --- negative above the edge, further suppressing the Fano factor as in Fig.~\ref{fig:fig1}e, but also positive below the edge in the purely quantum tunneling regime, where it counteracts the classical term by increasing the Fano factor to potentially super-Poissonian values ($F>1$). While the noise suppression can be achieved by memory of any origin, quantum or classical, the noise enhancement is a fingerprint of subtle quantum correlations.   

We now demonstrate these concepts by more elaborate analysis of the experimental data acquired at various values of the temperature and magnetic field. We start by fitting the experimental data for the mean current (insensitive to memory) around the edge(s) with a straightforward extension of the above theory to the case of two spin-split levels due to the magnetic-field with resulting 3x3 (double occupancy excluded) memory kernel analogous to equation \eqref{kernel} as shown in Fig.~\ref{fig:fig2}. In the finite magnetic field case 6 free parameters were fixed by fitting simultaneously curves at various temperatures, namely two independent critical exponents $\alpha_{\uparrow}=0.40$, $\alpha_{\downarrow}=0.43$ and thresholds $V_{\rm th \uparrow,\downarrow}$ together with an overall prefactor to the emitter rates and the leverage factor $\eta$, while in the $B=0$ case only 4 parameters due to a single resonance peak were fixed with $\alpha=0.28$ in qualitative agreement with in-depth investigations.\cite{Hapke-Wurst:PRB00,Frahm:PRB06}

By this procedure all parameter values are fully determined and the predicted Fano factor curves in the lower panel of Fig.~\ref{fig:fig3}
are free of any ambiguity. Considering this, the correspondence between the measurements (points with error-bars) and our non-Markovian theory (lines)
is quite remarkable in all cases encompassing two magnetic field values and various temperatures.
We also compare the Markovian, i.e., with the derivative terms in equation \eqref{Fanofactor} omitted (dashed lines), and non-Markovian (solid lines) predictions in the insets and the detail of Fig.~\ref{fig:fig3} with clear demonstration of the already mentioned non-Markovian features in the low-temperature Fano factor,
namely, the significantly more pronounced dip on the high-voltage side of the FES and the potentially super-Poissonian
peak on the low-voltage side with fast destruction of the non-Markovian corrections with temperature or distance from the resonant edge.

All these features are clearly seen in the experimental data as well. The super-Poissonian Fano factor due to quantum coherence at the lower edge is not reliably confirmed experimentally because of associated large errors resulting from a ratio of very small values of both the current and noise (tunneling regime). Nevertheless, the experimentally observed peak just below the upper edge (see the detail in Fig.~\ref{fig:fig3}), although sub-Poissonian, is caused by the very same mechanism and is thus an indirect confirmation of the purely quantum memory effect.  Altogether, the importance of the non-Markovian corrections due to quantum memory is established both qualitatively and quantitatively.
\newline

{\noindent\large\bf Methods\\}
{\small{\bf Experimental details.} The studied InAs QDs are embedded in a GaAs-AlAs-GaAs resonant tunneling device patterned into pillars with a cross section of $9\times9\unit{\mu m}$ sufficiently small to resolve single dot tunneling.\cite{Maire:PRB07} The effective AlAs barrier widths of 4 and 3\unit{nm} are slightly asymmetric. The measurements, whose schematic of the electronic setup is shown in Fig.~\ref{fig:fig1}a, were performed in a dilution refrigerator at temperatures down to $70\unit{mK}$ and magnetic fields up to $13\unit{T}$. The DC-part of the source drain current $I$ is measured with a transimpedance amplifier which also biases the sample. Two 4.7\unit{k\Omega} resistors convert the fluctuating current to voltages which are measured in a cross-correlation configuration. Together with parasitic capacitances these resistors form RC-circuits which define the bandwidth of our experiment. To increase this bandwidth we use home-built coaxial cables thereby lowering the total parasitic capacitance to $20\unit{pF}$.

The voltage fluctuations are amplified by a two-stage low temperature amplifier based on the ATF34143 HEMT with a gain of $22\unit{dB}$, followed by a room temperature amplifier with a gain of $60\unit{dB}$. The amplified signal is filtered and digitised, and from the Fourier spectra the cross-correlation noise power $S_{AB}$ is calculated and averaged over 8 minutes. To retrieve the shot noise power the real part of $S_{AB}$ in the frequency range from $500\unit{kHz}$ to $3\unit{MHz}$ is evaluated. A technical background-noise, largely dominated by thermal noise sources like the first transistor stage of the cryogenic amplifiers and the conversion resistors, is measured at zero current and subtracted. At finite sample impedances the partially correlated thermal background is estimated and also subtracted. The correlation gain parameters are determined by noise thermometry. The error bars in Fig.~\ref{fig:fig1}e and Fig.~\ref{fig:fig3} consist of the statistical error, the error of the estimated background and the calibration error.

{\bf Theoretical details} Hamiltonian of simplified spin-less model  of a resonant level tunnel-coupled to two leads (emitter E and collector C) and Coulomb-coupled just to the emitter reads $H=H_{QD}+H_{E}+H_{C}+H_{T}+V_{X}d^{\dagger}d$ with
$H_{QD}=\varepsilon_D d^{\dagger} d$,
$H_{\beta }=\sum_{k_{\beta}}\epsilon_{k_{\beta}}c_{k_{\beta}}^{\dagger}c_{k_{\beta}}$,
$H_{\beta T}=\sum_{k_{\beta}}(t_{k_{\beta}}d^{\dagger}c_{k_{\beta}}+t_{k_{\beta}}^{*}c_{k_{\beta}}^{\dagger}d)$ and
$V_{X}=\sum_{k_{E},k'_{E}}V_{k_{E},k'_{E}}c_{k_{E}}^{\dagger}c_{k'_{E}}$,
where $\beta=E,C$; $d$ and $c_{k,\beta}$ are QD and lead annihilation operators,
$\varepsilon_D$ and $\epsilon_{k_{\beta}}$ are the energies of the QD level and of the electrons in the leads,
respectively, while $t_{k_{\beta}}$ describe the tunneling between the QD and the leads.
The last term describes the scattering of emitter lead electrons on an electron in the QD and is responsible for the FES phenomenon. Since the Fermi level of the collector lead is far below the resonant level and $\gamma_{C}$ does not depend on energy close to the edge
(Fig.~\ref{fig:fig1}d) we can use the method by Gurvitz and Prager \cite{Gurvitz:PRB96}
to exactly integrate out the collector lead. This leads to the equation of motion for the density operator $\sigma(t;n)$ of the dot and
the emitter resolved with respect to the number $n$ of passed electrons through the emitter/QD interface.  After introducing the counting
field $\chi$ as a conjugate variable to $n$, one can write the equation of motion for $\mathbf{\sigma}(t;\chi)=\sum_{n}\mathbf{\sigma}(t;n) e^{n\chi}$, partly expressed in the block form,\cite{Flindt:PRL08, Flindt:PRB10}
$\mathbf{\sigma}=(\sigma_{00},\sigma_{11},\sigma_{01},\sigma_{10})^{T}$,
\begin{equation}\label{evolution}
\begin{split}
\frac{d\sigma(t;\chi)}{dt}&=-i\Big(H_E(\chi) \sigma(t;\chi)-\sigma(t;\chi)H_{E}(-\chi)\Big)\\
&+
\begin{pmatrix}
0&\gamma_C&0&0\\
0&-\gamma_C&0&0\\
0&0&-\gamma_C/2&0\\
0&0&0&-\gamma_C/2
\end{pmatrix}\sigma(t;\chi),
\end{split}
\end{equation}
where
$H_E(\chi)=H_{QD}+H_{E}+V_{X}d^{\dagger}d+\sum_{k_{E}}(t_{k_{E}} e^{\chi/2}d^{\dagger}c_{k_{E}}+t_{k_{E}}^{*} e^{-\chi/2} c_{k_{E}}^{\dagger}d)=H_{E}^{\dagger}(-\chi)$
is the appropriately modified Hamiltonian of the emitter and QD including the counting field.

The emitter lead can then be handled perturbatively in the tunnel coupling $t_{k_E}$ following closely the derivation for dissipative double quantum dot from Ref.~\onlinecite{Flindt:PRB10} by first separating equation \eqref{evolution} into four equations for the elements of $\sigma$.
Tracing out the electron states of the emitter lead in the two equations
for the evolution of the diagonal elements $\sigma_{jj}$ ($j=0,1$),
allows us to find the evolution equations for the generalised QD occupations $p_{j}(t;\chi)=\mathrm{Tr}_{E}\sigma_{jj}(t;\chi)$\label{rho11}
\begin{equation}\label{rho11}
\begin{split}
\frac{d p_{0/1}(t;\chi)}{dt}&=\pm \gamma_C p_1(t;\chi)\\
&\mp 2e^{\pm \chi/2}\sum_{k_{E}}\im\left[t_{k_{E}}\tr_E\left(c_{k_{E}}\sigma_{01}(t;\chi)\right)\right].
\end{split}
\end{equation}
The equation governing the evolution of $\sigma_{01}(t;\chi)$ which enters Eqs.~(\ref{rho11}) contains the diagonal elements $\sigma_{jj}(t;\chi)$ of the total density matrix. In order to close the equations for $p_{j}(t;\chi)$ we perform physically motivated QD-state-resolved perturbative decoupling of the density matrix into $\sigma_{jj}(t;\chi)=p_{j}(t;\chi)\otimes
\varrho^{E}_{j}$ with $\varrho^{E}_{j}=\exp\big(-[H_{E}+jV_{X}-\mu_{E}\sum_{k_{E}}c_{k_{E}}^{\dagger}c_{k_{E}}]/k_{B}T\big)/Z^{E}_{j}$ being the grand-canonical density matrix of the emitter lead at temperature $T$ and chemical potential $\mu_{E}$ when the QD is empty ($j=0$) or occupied ($j=1$). Thus, after multiplying the forward rate by 2 due to the interplay of spin and Coulomb blockade,\cite{Gurvitz:PRB96} we find equation \eqref{kernel} (reducing to equation \eqref{GME} for $\chi=0$) with FES Green's functions reading
$G_{0}(t)=\sum_{k_{E},k'_{E}}t_{k_{E}}t^{*}_{k'_{E}}\mathrm{Tr}_E\left[c_{k'_{E}}^{\dagger}e^{i(H_{E}+V_{X})t}c_{k_{E}}e^{-iH_{E}t}\varrho^{E}_{0}\right]$, $G_{1}(t)=\sum_{k_{E},k'_{E}}t_{k_{E}}t^{*}_{k'_{E}}\mathrm{Tr}_E\left[e^{i(H_{E}+V_{X})t}c_{k_{E}}e^{-iH_{E}t}c_{k'_{E}}^{\dagger}\varrho^{E}_{1}\right] $.
}


{\noindent\large\bf Acknowledgements\\}
We thank T.~L\"udtke and K.~Pierz for device fabrication, C.~v.~Zobeltitz for valuable discussions, and R.~Filip for useful comments on the manuscript.
This work was supported by the German Excellence Initiative via QUEST (Hannover),
by the Czech Science Foundation via Grant  No.~204/11/J042 (T.~N.), and the
TEAM programme of the Foundation for Polish Science, co-financed from the European Regional Development Fund (K.~R.).
\newline\newline
{\noindent\large\bf Author contributions\\}
The experiment was designed by F.H. and R.J.H. and carried out mostly by N.U. and N.M. Theory was designed and developed by K.R and T.N. N.U. and K.R. jointly analyzed and interpreted the data. Manuscript was written mainly by T.N. with contributions from N.U. and K.R. and with steady input and feedback from F.H. and R.J.H.    
\newline\newline
{\noindent\large\bf Additional information\\}
{\bf Competing financial interests:} The authors declare that they have no competing financial interests.


\begin{thebibliography}{31}

\bibitem{PhysicsToday}
Beenakker, C. \& Sch\"{o}nenberger, C. Quantum Shot Noise.
{\em Physics Today} {\bf 56,} 37--42 (2003).

\bibitem{Nazarov:book}
Nazarov, Yu.~V. {\em Quantum Noise in Mesoscopic Physics} (Springer, Berlin, 2003).

\bibitem{Blanter}
Blanter, Ya.~M. \& B{\"u}ttiker, M. Shot noise in mesoscopic conductors.
{\em Physics Reports} {\bf 336,} 1--166 (2000).

\bibitem{Glattli:NatPhys09}
Delattre, T. {\em et al.} Noisy Kondo impurities.
{\em Nature Physics} {\bf 5,} 208--212 (2009).

\bibitem{Yamauchi:PRL11}
Yamauchi, Y. {\em et al.} Evolution of the Kondo Effect in a Quantum Dot Probed by Shot Noise.
{\em Phys. Rev. Lett.} {\bf 106,} 176601 (2011).

\bibitem{Flindt:PRL08}
Flindt, C., Novotn\'y, T., Braggio, A., Sassetti, M.,
\& Jauho A.-P. Counting Statistics of Non-Markovian Quantum Stochastic Processes.
{\em Phys. Rev. Lett.} {\bf 100,} 150601 (2008).

\bibitem{Fleming:Science07}
Lee, H., Cheng, Y.-C., Fleming, \& G.~R. Coherence Dynamics in Photosynthesis: Protein Protection of Excitonic Coherence.
{\em Science} {\bf 316,} 1462--1465 (2007).

\bibitem{Erez:Nature08}
Erez, N., Gordon, G., Nest, M., \& Kurizki, G. Thermodynamic control by frequent quantum measurements.
{\em Nature} {\bf 452,} 724--727 (2008).

\bibitem{Eisert:PRL08}
Wolf, M.~M., Eisert, J., Cubitt, T.~S., \& Cirac, J.~I. Assessing Non-Markovian Quantum Dynamics.
{\em Phys. Rev. Lett.} {\bf 101,} 150402 (2008).

\bibitem{Breuer:PRL09}
Breuer, H.~P., Laine, E.~M., \& Piilo, J. 
Measure for the Degree of Non-Markovian Behavior of Quantum Processes in Open Systems.
{\em Phys. Rev. Lett.} {\bf 103,} 210401 (2009).

\bibitem{Plenio:PRL10}
Rivas,  \'A., Huelga, S.~F., \& Plenio, M.~B. Entanglement and Non-Markovianity of Quantum Evolutions.
{\em Phys. Rev. Lett.} {\bf 105,} 050403 (2010).

\bibitem{Piilo:NatPhys11}
Liu, B.~H. {\em et al.} Experimental control of the transition from Markovian to non-Markovian dynamics of open quantum systems.
{\em Nature Physics} {\bf 7,} 931--934 (2011). 

\bibitem{Hanggi:PRL05}
Mokshin, A.~V., Yulmetyev, R.~M., \& H\"anggi, P.
Simple Measure of Memory for Dynamical Processes Described by a Generalized Langevin Equation.
{\em Phys. Rev. Lett.} {\bf 95,} 200601 (2005).

\bibitem{Kubala:PRB06}
K\"onemann, J., Kubala, B., K\"onig, J., \& Haug, R.~J.
Tunneling resonances in quantum dots: Coulomb interaction modifies the width.
{\em Phys. Rev. B} {\bf 73,} 033313 (2006).

\bibitem{Rabani:PRB11}
Cohen, G. \& Rabani, E. Memory effects in nonequilibrium quantum impurity models.
{\em Phys. Rev. B} {\bf 84,} 075150 (2011).

\bibitem{Maire:PRB07}
Maire, N., Hohls, F., L\"udtke, T., Pierz, K., \& Haug, R.~J. 
Noise at a Fermi-edge singularity in self-assembled InAs quantum
dots.
{\em Phys. Rev. B} {\bf 75,} 233304 (2007).

\bibitem{Matveev:PRB92}
Matveev, K.~A. \& Larkin, A.~I. Interaction-induced threshold singularities in tunneling via
localized levels.
{\em Phys. Rev. B} {\bf 46,} 15337 (1992).

\bibitem{Geim:PRL94}
Geim, A.~K. {\em et al.} Fermi-edge singularity in resonant tunneling.
{\em Phys. Rev. Lett.}, {\bf 72,} 2061 (1994).

\bibitem{Hapke-Wurst:PRB00}
Hapke-Wurst, I. {\em et al.} Magnetic-field-induced singularities in spin-dependent tunneling
through InAs quantum dots.
{\em Phys. Rev. B} {\bf 62,} 12621 (2000).

\bibitem{Frahm:PRB06}
Frahm, H., von Zobeltitz, C., Maire, N., \& Haug, R.~J. Fermi-edge singularities in transport through quantum dots.
{\em Phys. Rev. B} {\bf 74,} 035329 (2006).

\bibitem{Nozieres:PR69}
Nozi\`eres, P. \& De~Dominicis, C.~T.  
Singularities in the X-Ray Absorption and Emission of Metals. III. One-Body Theory Exact Solution.
{\em Phys. Rev.} {\bf 178,} 1097--1107 (1969).

\bibitem{Ohtaka:RMP90}
Ohtaka, K. \& Tanabe, Y.
Theory of the soft-x-ray edge problem in simple metals: historical
  survey and recent developments.
{\em Rev. Mod. Phys.} {\bf 62,} 929--991 (1990).

\bibitem{Mahan:PR67}
Mahan, G.~D.  
Excitons in Metals: Infinite Hole Mass.
{\em Phys. Rev. } {\bf 163,} 612--617 (1967).

\bibitem{Andrei:PRL06}
Mehta, P. \& Andrei, N.
Nonequilibrium Transport in Quantum Impurity Models: The Bethe Ansatz for Open Systems.
{\em Phys. Rev. Lett.} {\bf 96,} 216802 (2006).

\bibitem{Doyon:PRL07}
Doyon, B. 
New Method for Studying Steady States in Quantum Impurity Problems: The Interacting Resonant Level Model.
{\em Phys. Rev. Lett.} {\bf 99,} 076806 (2007).

\bibitem{Schmitteckert:PRL08}
Boulat, E., Saleur, H., \& Schmitteckert, P. 
Twofold Advance in the Theoretical Understanding of Far-From-Equilibrium Properties of Interacting Nanostructures.
{\em Phys. Rev. Lett.} {\bf 101,} 140601 (2008).

\bibitem{Schmitteckert:PRL10}
Bransch\"adel, A., Boulat, E., Saleur, H., \& Schmitteckert, P.
Shot Noise in the Self-Dual Interacting Resonant Level Model.
{\em Phys. Rev. Lett.} {\bf 105,} 146805 (2010).

\bibitem{Gurvitz:PRB96}
Gurvitz, S.~A. \& Prager, Ya.~S. 
Microscopic derivation of rate equations for quantum transport.
{\em Phys. Rev. B} {\bf 53,} 15932 (1996).

\bibitem{Nauen:PRB04}
Nauen, A., Hohls, F., Maire, N., Pierz, K., \& Haug, R.~J.
Shot noise in tunneling through a single quantum dot.
{\em Phys. Rev. B} {\bf 70,} 033305 (2004).

\bibitem{Levitov:PRL05}
Abanin, D.~A. \& Levitov, L.~S.
Fermi-Edge Resonance and Tunneling in Nonequilibrium Electron Gas.
{\em Phys. Rev. Lett.} {\bf 94,} 186803 (2005).

\bibitem{Flindt:PRB10}
Flindt, C., Novotn\'y, T., Braggio, A., \& Jauho, A.-P. 
Counting statistics of transport through Coulomb blockade nanostructures: High-order cumulants and non-Markovian effects.
{\em Phys. Rev. B} {\bf 82,} 155407 (2010).

\end{thebibliography}
\end{document}